\documentclass[twocolumn,twoside,slac_two]{revtex4}
\usepackage{graphicx}
\usepackage{fancyhdr}
\begin{document}

\title{The Bayesian Block Algorithm}

%
\author{Jeffrey D. Scargle}
\affiliation{Space Science Division,
NASA Ames Research Center,
Moffett Field, California, 94035, USA
}
\author{Jay P. Norris}
\affiliation{Physics Department, Boise State University,
2110 University Drive,
Boise, Idaho, 83725, USA}

\author{Brad Jackson}
\affiliation{
San Jose State University, Department of Mathematics,
The Center for Applied Mathematics and Computer Science,
One Washington Square, MH 308,
San Jose, CA 95192, USA
}

\author{James Chiang}
\affiliation{SLAC National Accelerator Laboratory,
2575 Sand Hill Road, M/S 29,
Menlo Park, CA  94025, USA}


\begin{abstract}
This presentation briefly 
describes the \emph{Bayesian Block algorithm}
in the context of its application to analysis of 
time series data from the Fermi Gamma Ray Space Telescope.
Details are to be found 
in a recently published paper.
\end{abstract}

\maketitle

\thispagestyle{fancy}


\section{Introduction}

Much astrophysical information 
obtained with the Fermi Gamma Ray Space Telescope 
arises from analysis of time series data.  
Accordingly some of the desired capabilities
of science analysis software are the ability to:
\begin{itemize}

\item Detect
transient events, such as Gamma Ray Bursts, and flares in active galactic nuclei and 
other variable objects such as the Crab Nebula.

\item Characterize these events by modeling the shape of the light curve.

\item Estimate values of 
parameters from this characterization, such as rise and decay times, time scales of variability, and an index of variability.

\item Identify times over which the near constancy of the flux suggests good time intervals over which to carry out estimates of the energy spectrum

\end{itemize}

\section{The Bayesian Block Algorithm}

The Bayesian Blocks algorithm provides a useful 
method to address these problems.
The concept underlying  this approach is to find 
the best possible representation of 
time-sequential data as a series of segments
over which the underlying signal is constant
to within the observational errors.
The paper \citep{scargle_et_al}
describes a dynamic programming 
algorithm which yields such an optimal
price-wise constant model for time series data
in general, and time-tagged photon data
as a special case.

All operations of Bayesian Blocks can be implemented with no limitation on sampling, time resolution or signal amplitude.  
Data gaps and variable exposure are easily accommodated.  Other applications include data-adaptive histograms, multi-variate time series analysis and a novel approach to delay estimation in strong lensing events.

\section{Published Code}

Details and principles are given in \citep{scargle_et_al}.
This reference includes data files and 
Matlab\textregistered  \ code for implementing the Bayesian Blocks algorithm and reproducing the figures in the paper, 
thus implementing the discipline of Reproducible Research
as invented in
\citep{claerbout} and
devleoped in \citep{donoho_rr}.

Here are a few corrections to this paper and
the attached code:
\begin{itemize}

\item In the text of the Astrophysical Journal paper
\citep{scargle_et_al}
the natural logarithm symbol was omitted from
Equation 21, which should be
$$\mbox{ncp}\_\mbox{prior} = 4 - 
\mbox{log} (73.53  p_{0} N^{-.478} ) $$
This error was reported by Peter Williams, at
the Center for Astrophysics, Harvard University.
The attached code implements this 
expression correctly.

\item On line 222 of MatLab script named 
\verb+find_blocks.m+ the expression 
\verb+num_this = sum( nn_vec( ii_2: ii_2 ) );+
should be\\
\verb+num_this = sum( nn_vec( ii_1: ii_2 ) );+\\

That is, the summation of the populations of all of the data cells 
in the block is over the index range 
from 
\verb+ii_1+ to \verb+ii_2.+
This error,
only affecting the display of the block model
and not its computation, 
was reported by Mike Wheatland, the University of Sydney.

\item
In the same code package, 
the first line of the function script
\verb+plot_blocks_meas.m+
had a typo in the function name.
It should read\\
\verb+function rate_vec = plot_blocks_meas(+
\verb+  cpt_times, tt_vec, xx_vec, er_vec )+.
In most implementations of MatLab 
this error causes no problems.
\end{itemize}

Also it should be noted 
that this MatLab code was developed
with an earlier version (7.9, from 2009)
and there may be minor compatibility
issues with more recent versions.
As 
additional updates or corrections to this material are developed they will be posted at\\
\verb+http://bayesianblocks.blogspot.com+

In addition Jake Vanderplas'Êrelated blog 
``Dynamic Programming in Python: Bayesian Blocks'' 
nicely described the algorithmic approach of dynamic programming, with examples in the context of histograms:
\verb+http://jakevdp.github.io/blog/+
\verb+2012/09/12/dynamic-programming-in-python/+.
See also the posting at the Starship Asterisk discussion forum:
\verb+ http://asterisk.apod.com/+
\verb+viewtopic.php?f=35&t=29458+

\bigskip 
\begin{acknowledgments}
This work has been supported by the NASA
Applied Information Sciences Research Program;
we are particularly indebted to Joe Bredekamp
for his leadership and encouragement.
Additional support was provided by the Woodward Fund,
the Center for Applied Mathematics and Computer Science,
San Jose State University, and we gratefully acknowledge
the contributions of the many students in this program.
We are also indebted to  
Jake Vanderplas,
Peter Williams
and Mike Wheatland,
for helpful suggestions, 
and to Alice Allen
for assistance in setting up
the posting for the 
Starship Asterisk discussion forum.

\end{acknowledgments}

\bigskip 

\begin{thebibliography}{9}   


\bibitem[Claerbout(1990)]{claerbout}
J. Claerbout,
``Active documents and reproducible results,''
Stanford Exploration Project Report, Vol. 67, p. 139, 1990

\bibitem[Donoho et al.(2008)]{donoho_rr}
D. Donoho,  A. Maleki, I. Rahman,
M. Shahram and V. Stodden, 
``15 Years of Reproducible Research in
Computational Harmonic Analysis,''
Computing in Science and Engineering, 11, 8, 2009,
\verb+http://stats.stanford.edu/~donoho/Reports/+\\
\verb+2008/15YrsReproResch-20080426.pdf+


\bibitem[Scargle et al.(2013)]{scargle_et_al}
J. D. Scargle, J. P. Norris, B. Jackson, and J. Chiang, 
 ``Studies in Astronomical Time Series Analysis. VI. Bayesian Block Representations,'' Astrophysical Journal, Vol. 764, p. 167, 2013,
 arXiv: 1207.5578.




\end{thebibliography}

\end{document}